\documentclass[11pt,superscriptaddress]{article}                        
\usepackage{amsmath}
\usepackage{graphicx}

\usepackage{xcolor} 
\usepackage[normalem]{ulem}

\newcommand{\ber}{\begin{eqnarray}}
\newcommand{\eer}{\end{eqnarray}}
\newcommand{\bea}{\begin{equation}}
\newcommand{\eea}{\end{equation}}
\newcommand{\del}{\partial}
\protected\def\verythinspace{%
  \ifmmode
    \mskip0.3\thinmuskip
  \else
    \ifhmode
      \kern0.050004em
    \fi
  \fi
}

\begin{document}
\title{Path integral estimates of the quantum fluctuations of the relative soliton-soliton velocity in a Gross-Pitaevskii breather} 

\author {\bf Sumita Datta \\
Alliance School of Applied Mathematics\\
Alliance University\\
Bengaluru 562 106, Karnataka, India\\
\bf Vanja Dunjko \\
\bf Maxim Olshanii \\
Department of Physics \\
University of Massachusetts Boston\\
Boston MA 02125, USA} 

\maketitle

\begin{abstract}
We study the quantum fluctuations of the relative velocity of constituent solitons in a Gross-Pitaevskii breather. The breather is confined in a weak harmonic trap. These fluctuations are monitored, indirectly, using a two-body correlation function measured at a quarter of the harmonic period after the breather creation. We compare the results of an ab initio
% "Foreign words do not need to be highlighted or italicized, including Latin terms such as ‘in situ’." (https://www.mdpi.com/authors/layout)
quantum Monte Carlo calculation based on the Feynman-Kac path integration method with the analytical predictions founded on the Bogoliubov approximation [O. V. Marchukov {\it et al}, PRL {\bf 125}, 050405 (2020)] and find a satisfactory agreement.

\end{abstract}
\section{Introduction}
A purely solitonic solution the focusing nonlinear Schr\"{o}dinger equation (NLSE) consists of a finite number of solitons, each parameterized by its norm, velocity, initial position, and initial phase. A single stationary soliton is known as the fundamental soliton. When a solution consists of two or more solitons that are at rest relative to each other and have the same initial positions, it is commonly known as a \emph{breather}. The name comes from the fact that the density profile of this kind of solution periodically oscillates in time (provided the constituent solitons have unequal amplitudes). Certain kinds of breathers can be produced by quenching the strength of the nonlinear interaction. Mathematically, this means the following: let $\psi(x)$ be the fundamental soliton of an NLSE, at some point of time. Suppose we use this $\psi(x)$ as the initial condition for time propagation under an NLSE whose interaction strength is four times that of the original NLSE. We will find that the propagated solution is a 2-breather whose constituent solitons have norms that are in a ratio of 3:1. Experimentally, this means that a sudden increase of the interaction strength by a factor of 4 converts a fundamental soliton to a 3:1 breather. This was analytically predicted by Satsuma and Yajima \cite{2} in 1974 and was recently experimentally verified in dilute Bose-Einstein condensates (BECs) \cite{3,4}. Now, in experiments, there is always a background trapping potential, so that the relevant equation for BEC experiments is the NLSE plus external harmonic confinement. The resulting equation is called the 1D Gross-Pitaevskii equation (GPE). Corresponding to the NLSE breathers, there are Gross-Pitaevskii breathers (GP breathers). In this paper, we compute a correlation function associated with a GP breather corresponding to the 3:1 NLSE breather, using a path integral technique.

Yurovsky et al.\ \cite{5} showed that quantum many-body effects cause an NLSE breather to disassociate into its constituent solitons. Recall that the 3:1 breather consists of two solitons of norms $N/4$ and $3N/4$ (where $N$ is the number of atoms in the condensate), which in the mean-field approximation (i.e.\ at the level of the GPE) sit on the top of each other and do not move. Using the Bethe ansatz, Yurovsky and coworkers found that quantization leads to a drift in the relative position of the constituent solitons. Unfortunately, this method becomes unusable for more than $N=23$ atoms, and the prediction for experimentally relevant, larger $N$ was an extrapolation. Nevertheless, it was predicted that the relative soliton-soliton velocity in a GP breather may become experimentally observable, for empirically realistic propagation times. Further work corroborated this prediction using two different kinds of approximation: \cite{1} used the Bogoliubov approach,  while \cite{7} used the truncated Wigner approximation. The predicted quantum fluctuations of macroscopic variables are yet to be observed, but the breather itself is already an experimental reality  \cite{3,4}. On the theoretical side, however, we see that there is still a need for a nonpertubative, fully quantum-mechanical approach for building a quantum counterpart of the GP breather, one that is usable for a number of particles $N$ that is substantially larger than 23. 

%As the present numerical investigation shows, such an approach is a quantum Monte Carlo method based on the Feynman-Kac (FK) path integration% \cite{6}, which we use to estimate the quantum fluctuations of the relative soliton-soliton velocity in a GP breather. We present an ab init%io confirmation of the observability of the quantum fluctuations of the relative soliton-soliton velocities, using the path integral Monte Ca%rlo method \cite{6}. 
In order to estimate the quantum fluctuations of the relative soliton-soliton velocity in a GP breather, we adopt a quntum Monte Carlo method
based on the Feynman-Kac path integration[6]. An ab initio confirmation of the observability of the quantum fluctuations of the relative soliton-soliton velocities is presented here using the path integral method[6].
Metropolis \cite{8} was the first to exploit a relationship between the Schr\"{o}dinger equation for imaginary time and the random-walk solution of the general diffusion equation. We are considering the initial-value problem
\ber
& & \frac{\del {\psi}(x,t)}{\del t}=(\frac{\Delta}{2}-V){\psi}(x,t)\\
%\nonumber
%\\
& & {\psi}(x,0)=f(x)\,, \nonumber \\
\,\,,
\nonumber
\eer
with $x \in  R^d$ and $\psi(x,0)=f(X)$. The solution of Equation (1) 
% it seems that this journal prefers not to abbreviate 'Equation'
can be written in Feynman-Kac representation as
\bea
\psi(x,t)=E_x[e^{-\int_0^t V(X(s))ds}f(X(t)]\,,
\eea
where $X(t)$ is a Brownian motion trajectory, $E_x$ is the average value of the exponential term with respect to these trajectories, and $f$ is the initial value of $\psi$, the latter being the sought-after solution of the above Cauchy problem. This classical representation of the time-dependent solution to the Schr\"{o}dinger equation involves a Wiener measure \cite{9} (i.e., the probability measure on the space of continuous functions) and, unlike the ordinary path integration, provides a rigorous mathematical justification. The above representation is used to calculate the energies and any correlation properties associated with any particular solution. It is straightforward to implement numerically and does not require a trial function. This method was first applied to calculate energy eigenvalues of simple systems by Donsker and Kac \cite{10} and was eventually extended to atomic systems by Korzeniowski et al.\ \cite{11}. It is also known that the classical FK formula provides a rather slow rate of convergence due to the fact that the underlying diffusion process---Brownian motion (Wiener process)---is non-recurrent. Specifically, in dimensions higher than two, the trajectories of the process escape to infinity \cite{12} with probability one. Mathematically speaking, for $d \ge 3$, $P[\displaystyle \lim_{t\to\infty} R_t=\infty]=1$ (where $R_t$ is the distance of the Brownian particle from the origin) and $\displaystyle \lim_{t\to\infty} P(X(t)\in B)=0$. Here $B$ is a Borel set, the set of all Brownian processes $X(t)$. As a result, the sampling within the quantum-mechanical region of intersection occurs only during a small fraction of the total simulation time and the rate of convergence becomes prohibitively slow. Another path integral method, known as the generalized Feynman-Kac method, was initiated by Soto and Claverie \cite{13}  and was subsequently extended to the full generalized Feynman-Kac (GFK) method by Caffarel and Claverie \cite{14}. These procedures can be considered an application of importance sampling to the FK integral, along with the transformation of Equation (1) into a Wiener path integral over a distribution determined by both the diffusion and drift terms. This transforms the difficult-to-handle branching (potential energy) term into a more manageable path integral. The GFK method is mathematically more convenient because the limiting distribution exists, i.e., $\displaystyle \lim_{t\to\infty} P(X(t)\in B)=\int_B{{\phi}_T}^2(x)dx$. We use the GFK method to calculate the two-body correlation functions as a measure of the quantum fluctuation of the soliton-soliton relative velocity in a Gross-Pitaevskii breather. We find that our numerical estimate for the quantum fluctuations compares favorably with the preliminary theoretical estimates and is consistent with the Bogoliubov prediction \cite{1}.

\section{The model}

For calculating the two-body correlation function, we first consider the ground state of a solitonic system consisting of bosonic atoms (say, ${}^{7}\textrm{Li}$) with a negative scattering length. We assume that a 2-soliton breather is created at $t=0$ by quenching the interaction strength of the initial Hamiltonian.    

In what follows, we use the system of units in which
% don't use the double dollar sign; see https://tex.stackexchange.com/questions/503/why-is-preferable-to
\[
\hbar = m = \omega =1 
\,\,,
\]
where $m$ is the atomic mass and $\omega$ is the frequency of the harmonic confinement.  

%The Hamiltonian we are interested in is 
%\bea
%\hat{H}_{0}=-\frac{1}{2}\Delta+V(x)
%\label{H}
%\,\,,
%\eea
%where
%\ber
%&&
%\Delta = \sum_{i=1}^{N} \frac{{\partial}^2}{{\partial {x_i}^2}}
%\nonumber
%\\
%&&
%V=V_{\text{int}}+V_{\text{trap}}
%\nonumber
%\\
%&&
%V_{\text{int}}=-\tilde{g}\sum_{i=1}^{N-1}\sum_{j=i+1}^{N}{\delta}_{\tilde{\sigm%a}}(x_i-x_j)
%\label{V_int}
%\\
%&&
%V_{\text{trap}}= \frac{1}{2}\sum_{i} x_i^2
%\nonumber
%\,\,.
%\eer
%Here
%$$
%\tilde{g} \equiv \frac{|g|{\sqrt m}}{\hbar^{\frac{3}{2}}\sqrt{\omega}}
%$$
%is the dimensionless form of the absolute value of the one-dimensional coupling% constant, $g$ is the one-dimensional coupling constant, and $N$ is the number %of bosonic particles.

We would like to perform a time propagation of the time-dependent Schr\"{o}dinger equation 
\bea
i\frac{\partial}{\partial{t}}|\psi(t)\rangle=\hat{H_0}|\psi(t)\rangle
\label{time-dependent_schrodinger}
\,\,,
\eea
with the initial condition given by the ground state of the initial Hamiltonian
\bea
\hat{H}_{0}=-\frac{1}{2}\Delta+V_{0}(x)
\,\,,
\eea
where
\ber
\Delta = \sum_{i=1}^{N} \frac{{\partial}^2}{{\partial {x_i}^2}}
\eer
\ber
&&
V_{0}=V_{\text{int,\,0}}+V_{\text{trap}}
\nonumber
\\
&&
V_{\text{int,\,0}}=-\tilde{g}_0\sum_{i<j}{\delta}_{\tilde{\sigma}}(x_i-x_j)
\label{V_int_0}
\nonumber
\,\,,
\eer
\bea
V_{\text{trap}}= \frac{1}{2}\sum_{i} x_i^2\,.
\eea 
Here
\[
\tilde{g}_{0} \equiv \frac{|g_{0}| \sqrt{m}}{\hbar^{\frac{3}{2}}\sqrt{\omega}}
\]
is the dimensionless form of the absolute value of the initial coupling constant $g_{0}$ and $N$ is the number of bosonic particles. We assume that a 2-soliton breather is created at $t=0$. For that to happen, the pre- and post-quench values of the coupling constant must be related as   
\bea
g_{0} = \frac{1}{4} g
\,\,.
\eea
To estimate the quantum fluctuations of the relative velocity of the constituent solitons in a Gross-Pitaevskii breather, we adopt a path integral approach \cite{15,16} based on the Feynman-Kac integral formalism. To write the path integral solution, we first consider the Cauchy problem related to the time-dependent Schr\"odinger equation in Equation (3) 
\ber
& & i\frac{\del {\psi}(x,t)}{\del t}=(-\frac{\Delta}{2}+V){\psi}(x,t)
\nonumber
\\
& & {\psi}(x,0)=f(x)
\,\,,
\nonumber
\eer
 with a Hamiltonian
\ber
H=-\Delta/2+V(x)
\,\,.
\nonumber
\eer
The solution of the above equation in the Feynman-Kac representation can be written as 
\bea
  {\psi}(x,t)= E_x\{e^{-{\int^t_0 V(X(s))ds}}f(X(t))\}
   \eea
where $E_x$ is the expectation value of the random variables and $f$ is the initial value of the wavefunction $\psi$.  As was mentioned in the introduction, even though the FK formalism provides a basis for rigorous and accurate calculations of ground- and excited-state properties of many-particle systems, it suffers from a slow convergence rate due to the fact that the underlying diffusion process, Brownian motion (Wiener process), is non-recurrent. To speed up the convergence, one needs to use the Generalized Feynman-Kac (GFK) formalism, as described below. The GFK formalism employs an Ornstein-Uhlenbeck process $Y(t)$, which has a stationary distribution and the convergence becomes much faster. The solution in the Feynman-Kac representation holds for any potential $V$ which belongs to the Kato class \cite{17}. All the ordinary potentials fall under this category. One can obtain the GFK formalism from the raw Feynman-Kac representation by allowing a large class of diffusions that, unlike Brownian motion, have stationary distributions. Specifically, for any twice-differentiable positive $\phi(x)$, one defines a new potential $U$ as a perturbation of the potential $V$:
\bea
U(x)=V(x)-\frac{1}{2} \frac{\Delta \phi(x)}{\phi(x)}\,.
\eea
Then one has
\ber
\frac{\partial w(x,t)}{\partial t}
& & =\frac{1}{2}\Delta w(x,t)+\frac{\nabla \phi(x)}{\phi(x)}.\nabla w(x,t)-U(x)w(x,t)
\label{former_18}\\
& & =-Lw(x,t)
\nonumber\\
& & w(x,0)=g(x)\,,
\nonumber
\eer
where $g$ is the initial value of $w$. 
Equation (10) has the solution
\ber
w(x,t)=
E_x\{e^{-{\int^t_0 U(Y(s))ds}}g(Y(t))\}\,,
\label{former_21}
\eer
where the new diffusion $ Y(t) $ has an infinitesimal generator $A=\frac{\Delta}{2}+\frac{\nabla \phi}{\phi}\nabla $, whose adjoint is $A^{\star}(\cdot)=\frac{\Delta}{2}-\nabla(\frac{\nabla \phi}{\phi}(\cdot))$. Here ${\phi}^2(x)$ is a stationary density of $ Y(t) $, or equivalently, $A^{\star}({\phi}^2)=0$.

To see the connection between $ w(x,t) $ and $ u(x,t) $, observe that for $ f=1 $ and $g=1$,
\bea
w(x,t)=\frac{{\psi(x,t)}}{\phi(x)}\,,
\label{former_23}
\eea
because $ w(x,t) $ satisfies Equation (10). The diffusion $ Y(t) $ solves the following stochastic differential equation:
$dY(t)=\frac{\nabla \phi(Y(t))}{\phi(Y(t))}+dX(t)$\,.

%%%%%%%%%%%%%%%%%%%%%%%%%%%%%%%%%%%%%%%%%%%%%%%%%%%%%%%%%%%%

For numerical calculations, we will be using the Gaussian representation for the delta-function potential,
\ber 
{\delta}_{\tilde{\sigma}}(x_i-x_j)
=\frac{1}{\sqrt{2{\pi}}{\tilde{\sigma}}}e^{-\frac{(x_i-x_j)^2}{2{\tilde{\sigma}}^2}}\,.
\eer

For the purpose of the path-integral Monte Carlo, the system is represented by a $d$-dimensional particle, with $d=N$, subject to the boundary conditions 
\ber 
&&
\psi{\mid}_{\vec{x}={\pm}\infty}=0
\nonumber
\\
&&
\frac{\partial\psi}{\partial \vec{x}}{\mid}_{\vec{x}={\pm}\infty}=0
\nonumber
\,\,.
\eer
 
The raw Feynman-Kac formula \cite{6} will provide a Cauchy-type solution and we will adopt a guided random walk using a trial function that will satisfy the required boundary conditions. Using Equation~(10) and the GFK path integral representation  \cite{14}, the solution to Equation~(3) can be represented as \cite{15} 
\ber
\psi(x,t)
=w(x,t){\phi}(x)
={\phi}(x)E_x[e^{-U(Y(s))ds}]\,.
\label{former_8}
\eer
In Equation (14), there are two sums: the modified potential $U(Y(s))$ is summed over all the steps in a given trajectory, and then $[e^{-U(Y(s))ds}]$
is summed over all the trajectories.
Here in this computational problem a soliton can be viewed as a bound state of  ${}^{7}\textrm{Li}$  atoms interacting through a strong attractive potential described in Eq(13). The nonnegative function ${\phi}(x)$ can be chosen as a trial function consistent with the symmetry of the problem. In the present case we choose the following function for ${\phi}(x)$ and label it  as ${\phi_0}(x)$: 
\bea
{\phi}_0(x)=Ce^{-bx^2}\,,
\label{former_9}
\eea
where $C$ is normalization constant and $b$ is a variational parameter.
%\bea
%\text{\it density}=
%{\left|{\psi}\right|}^2
%\label{former_10}
%\eea
Now in terms of $e_0$ and $U(x)$ we introduce a new perturbed potential
\bea
V_p(x)=e_0-U(x)=
e_0-V(x)-\frac{1}{2} \frac{\Delta {\phi}_0(x)}{{\phi}_0(x)}
\eea
where $e_0$ is the energy associated with the trial function ${\phi}_0$.
The Ornstein-Uhlenbeck process $Y(t)$ is related to Brownian process $X(t)$ as follows: 
\bea
dY(t)=\frac{\nabla \phi(Y(t))}{\phi(Y(t))}+dX(t)
\eea
%In Equation (14), there are two sums: the modified potential $U(Y(s))$ is summed over all the steps in a given trajectory, and then $[e^{-U(Y%(s))ds}]$
%is summed over all the trajectories.
% '\text{\it npi}' represents such trajectories and total number of steps,  $\text{\it steps}\equiv \text{\it scale}^2 \times
%t$ (see Equations (11) and (14)).
In Eq(17) the first and second term represent the drift and diffusion respectively and the presence of these terms  
in this expression enables the trajectory $Y(t)$ to be highly localized. As a result, the important regions of the potential are frequently sampled and Equation (14) converges rapidly. Similarly, the expectation value of the operator $A$ is given by \cite{14}
\bea
{\langle Y|A|Y\rangle}=
\frac{\lim_{t\to\infty}\int dY(t)A(Y(t))e^{-\int[{V}_p(Y(s)ds}}
{\int dY(t)e^{-\int{V}_p(Y(s)ds}}\,.
\label{former_13}
\eea 
Equation~(18) is the key formula we will be using to calculate the quantum fluctuations of the soliton-soliton velocity. 

Our goal is to compute the second moment of the two-body relative distance, at $t=T/4$:
\bea
\langle\psi(t=T/4)|(x_1-x_2)^2|\psi(t=T/4)\rangle
\,\,.
\nonumber
\eea
The reason we are interested in it is that this particular correlator is sensitive to the quantum fluctuations of the relative soliton-soliton velocity, itself a macroscopic variable. 
The large-$N$ analytic predictions for the velocity fluctuations have been computed in \cite{1}:   
\bea
\langle\psi(t=0)|V_{\text{rel.}}^2|\psi(t=0)\rangle = 0.0429 \,\tilde{g}^2 N
\,\,, 
\nonumber
\eea
where the numerical prefactor comes from a numerically computed integral. (The formula  above neglects the zero-point quantum fluctuations that are induced by the harmonic confinement.) In turn, the variance of the relative distance between the centers of mass of the two solitons, after a quarter of a period, will be given by
\bea
\langle\psi(t=T/4)|X_{\text{rel.}}^2|\psi(t=T/4)\rangle = \langle\psi(t=0)|V_{\text{rel.}}^2|\psi(t=0)\rangle
\,\,. 
\nonumber
\eea 

Now, the mean occupations of the two constituent solitons are $N/4$ and $3N/4$. Therefore, the probability that two detected particles, 1 and 2, belong to two different solitons is $6/16$. Furthermore, assume that at $T/4$, the distance between the solitons exceeds their width, and therefore the 1 to 2 distance will be dominated by the distance between the centers of mass of the two solitons. Since $(6/16)\times 0.0429 = 0.0161$, we obtain
\bea
\langle\psi(t=T/4)|(x_1-x_2)^2|\psi(t=T/4)\rangle\approx 0.0161{\tilde{g}}^2 N 
\,\,.
\label{theory_prediction}
\eea

As we mentioned above, this estimate assumes that (a) the interaction-induced fluctuations in the relative velocity of the solitons exceed those generated by the zero-point fluctuations of the trap, and (b)  the soliton-soliton separation at the quarter of the period exceeds the size of the initial density distribution. Let us check the validity of these assumptions. For (a), we compare the separation \eqref{theory_prediction} with the zero-point fluctuations of the relative distance, $\sqrt{(8/3)/N}$. For (b), we consider the worst-case scenario and assume that the detected atoms 1 and 2 were found on the opposite wings of their respective solitons. That would require that the soliton-soliton distance given by \eqref{theory_prediction} exceeds the resulting correction to the distance, $\ell_{N/4} + \ell_{3N/4}$, where $\ell_{N'} \equiv 2/(\tilde{g}N')$ is the size of a soliton with $N'$ atoms. We get
\ber
& & 
\frac{1}{(\tilde{g}N)^2}  \ll 0.06
\label{neglect_zero-point}
\\ 
& & 
\frac{1}{(\tilde{g}N)^2}  \ll \frac{0.01}{\sqrt{N}}
\label{neglect_soliton_size}
\eer   
for the conditions (a) and (b), respectively.

The left-hand side of Equation~(19) will be numerically calculated using Equation~(18), where we will set $A=(x_1-x_2)^2$.
 Here $T\equiv 2\pi$ is the dimensionless form of the trapping period. 
\newpage

%%%%%%%%%%%%%%%%%%%%%%%%%%%%%%%%%%%%%%%%%%%%%%%%%%%%%
\section{Results}
The variance in the particle-particle distance that we are interested in is shown in Table 1. The agreement between the theoretical predictions and our numerical results is satisfactory, with the exception of the $\tilde{g}=0.78$ case. At the moment, condition \eqref{neglect_soliton_size} is barely satisfied, and it is likely that the solitons' width still contributes to the variance in the interparticle distance.
\begin{table}[h!]
\begin{center}
\caption {\bf Quantum fluctuations  calculated from theory and from numerical work.}
\label{t:results}
\vskip 0.5cm
\begin{tabular}{|c|c|c|c|c|c|c|c}
\hline
 $N$ & \text{\it scale} & \text{\it npi} & $\tilde{g}$ & $\tilde{\sigma}$ &  $\langle(x_i-x_j)^2\rangle $(Numerics)  & $ \langle(x_{i}-x_{j})^2\rangle $(Theory)\\
% Np & scale & \text{\it npi} & gT & sigT\\
\hline
100 & 30 & 50 & 0.5  & 0.016 & 0.5884$\pm 0.1647$ & 0.4025\\
    &    &    & 0.55 & 0.015 & 0.5309$\pm 0.1486$ & 0.487\\
    &    &    &  0.61 & 0.015 & 0.6209$\pm 0.1738$ & 0.599\\
    &    &    &  0.78 & 0.012 & 0.4389$\pm 0.1229$ & 0.9795\\
    &    &    &  0.83 & 0.01  & 1.6773$\pm 0.4696$ & 1.1091\\
    &    &    &  0.85 & 0.01  & 1.8155$\pm 0.5083$ & 1.632\\
\hline
\end{tabular}
\end{center}
\end{table}
\begin{table}[h!]
\begin{center}
\caption{\bf Notation Table}
\begin{tabular}{|c|c|c}
\hline
symbols &Physical Quantities \\
\hline
$N$  & Number of atoms \\
\hline
$scale$ & $\sqrt{(number  of  steps)/t}$ \\
\hline
$ t $ & simulation time \\
\hline 
\text{\it npi} & number of trajectories \\
\hline
$\tilde{g}$ & dinesionless form of the interaction strength \\
\hline
$\tilde{\sigma}$ & dimensionless form of the width of the Gaussian Potential  \\
\hline
$\epsilon$ & binomially distributed random variables\\
\hline
$B$ & Borel set\\
\hline
$C$ & Normalization constant in the trial function\\
\hline
\end{tabular}
\end{center}
%\vspace{-5 cm}
\end{table}

In order to make a connection with experiments with $^{7}\text{Li}$  \cite{4}, we assume that the mass of an individual atom is $m=7.016 \, \text{u}$ (where u is the unified atomic mass unit), 
% The SI brochure does not use 'amu'. Instead, it uses the dalton, symbol 'Da', or, alternatively, the  'unified atomic mass unit', symbol 'u'.
the post-quench scattering length $a_{\text{sc}}=-16.2 \, a_{0}$ (where $a_{0}$ is the Bohr radius), and the frequency of the radial trapping potential $\omega_{r}=2\pi \times 297 \, \text{Hz}$. The coupling constant is given by $g=2\hbar\verythinspace \omega_{r}\verythinspace a_{\text{sc}}$. This set reproduces  the conditions of the Rice experiment \cite{4} \emph{verbatim}, with the exception of the scattering length $a_{\text{sc}}$. The different value of $a_{\text{sc}}$ accounts for the difference between the number of atoms in the experiment and that in our work. In particular,  we adjusted $a_{\text{sc}}$ so that the ratio between the number of atoms and the condensate collapse threshold,  $N_{c} = 0.67 a_{\text{r}}/|a_{\text{sc}}|$ \cite{18}, is the same as it was in the experiment. The first line of the table above, $\tilde{g} = 0.55$, would correspond to a one-dimensional trapping frequency of $\omega={g^2 m}/({\tilde{g}^{2}\hbar^{3}})= 3.7\times 10^{-3}\,\text{Hz}$, corresponding to a propagation time of $T/4 = 4.3\times 10^{2} \,\text{s}$. 

This propagation time appears to be much longer than the estimate of $4.7\, \text{s}$ in \cite{1}, for $N = 3\times 10^{3}$. However, the former corresponds to a conservative estimate, based on a soliton-soliton separation exceeding six half-widths of the broader soliton. For a less conservative requirement used in \cite{1}, the propagation time will be as short as $T/4=0.83\, \text{s}$, reached with $\omega= 1.9\;\text{Hz}$ (accordingly, $\tilde{g} = 0.024$), with the rest of the experimental conditions kept intact.

%%%%%%%%%%%%%%%%%%%%%%%%%%%%%%%%%%%%%
\section {Conclusion and outlook}
We have confirmed that quantum fluctuations of a macroscopic observable---represented by the relative velocity of two solitons in a harmonically trapped Gross-Pitaevskii breather---can be observed in a predominantly mean-field environment. The scheme involves a harmonic quarter-period propagation of the breather: this time turns out to be sufficient for the breather to dissociate through purely quantum effects. As a computational method, we used the path integral Monte Carlo: our numerical results are consistent with the earlier predictions based on the Bogoliubov approximation \cite{1}.

In future work, will consider stronger interactions, both to move closer to the experimental  conditions and to suppress the residual effects of the soliton width. We expect that the future analog of Table 1 will exhibit a closer correspondence between the theory and numerics.
\newpage
%%%%%%%%%%%%%%%%%%%%%%%%%%%%%%%%%%%%%

%%%%%%%%%%%%%%%%%%%%%%%%%%%%%%%%%%%%%

\section{Appendix A: Numerical details} 
The formalism described in section 2 can include any generalized potential \cite{19} and is valid for any arbitrary dimension d (d=3N). To implement Equation (3) numerically, the 3N-dimensional Brownian motion can be replaced by properly scaled one-dimensional random walks as follows \cite{11,16,20}: 
\ber
W(l)\equiv W(t,n,l)
& = & {w_1}^1(t,n,l),{w_2}^1(t,n,l),{w_3}^1(t,n,l)....\\ \nonumber
&   &                  .......{w_1}^N(t,n,l){w_2}^N(t,n,l){w_3}^N(t,n,l)
\eer
where
\bea
{w_j}^i(t,n,l)=\sum^l_{k=1}\frac{{\epsilon}^i_{jk}}{\sqrt n}\,,
\eea
with ${w_j}^i(0,n,l)=0$ for $i=1,2,....,N$, $j=1,2,3$, and $l=1,2,.....,nt$. Here the $\epsilon$ variables denote the binomially distributed random variables which are 
chosen independently and randomly with probability $P$ for all $i,\,j,\,k$ such that
$P({\epsilon}^i_{jk}=1)$=$P({\epsilon}^i_{jk}=-1)$=$\frac{1}{2}$. It is known 
(from an invariance principle \cite{21}) that for every $\nu$ and $W(l)$
defined in Equation~(23),                                                                                
\ber
\lim_{n\to\infty}P(\frac{1}{n}\sum^{nt}_{l=1}V(W(l)))\leq \nu \\ \nonumber
 =  P( \int\limits^t_0 V( X(s))ds)\leq\nu
 \,\,.
\eer
Consequently, for large n,
\ber
P[ \exp(- \int\limits^t_0 V(X(s))ds)\leq\nu ] \\ \nonumber
 \approx  P [\exp(-\frac{1}{n}\sum^{nt}_{l=1}V(W(l)))\leq \nu]\,.
\eer
Finally, by generating $N_{rep}$ independent realization $Z_1$,$Z_2$,....$Z_{N_{rep}}$ of
\bea
Z_m=\exp(-(-\frac{1}{n}\sum^{nt}_{l=1}V(W(l)))
\eea
and using the law of large numbers, with regard to Equation~(24), we conclude that 
\bea 
(Z_1+Z_2+...Z_{N_{rep}})/N_{rep}=Z(t)
\eea
is an approximation of Equation~(8).
%\bea
%\mu\approx -\frac{1}{t}\log Z(t)
%\eea
Here $W^m(l)$, with $m=1,\,2,\,\ldots\,N_{rep}$, denotes the $m^{th}$ realization of $W(l)$ out of $N_{rep}$ independently run simulations. In the limit of large t and $N_{rep}$, this approximation approaches an equality and forms the basis of a computational scheme for the lowest energy of a many particle system with a prescribed symmetry.

\newpage
%%%%%%%%%%%%%%%%%%%%%%%%%%%%%%%%%%%%%

%%%%%%%%%%%%%%%%%%%%%%%%%%%%%%%%%%%%%

\section{Appendix B: Validity of the Gaussian approximation for the $\delta$-function}
To ensure that the pairwise Gaussian potential correctly describes the intended delta potential, one must check the following:
(i) the potential must support only one bound state, and
(ii) the energy of the bound state must be less than the potential depth.
To prove that the Gaussian potential with our choice of parameters supports
only one bound state we have followed the prescription given in reference  \cite{22}. In one dimension, the WKB integral for the energy E is given by  \cite{23} $ \int^{x_2}_{x_1}\sqrt {2[E-V(x)]} dx=(n-\frac{1}{2})\pi $,
where $x_1$ and $x_2$ are the turning points of the classical motion. 
For the Gaussian potential, the quantum number N of the last bound state can be obtained by using the above WKB formula for $E=0$: 
$\sqrt{2V_0}\int^{+\infty}_{-\infty}e^{-\alpha {x^2/2}}dx=(N-\frac{1}{2})\pi$, which gives $N=\frac{2}{\sqrt\pi}\sqrt{\frac{V_0}{\alpha}}+\frac{1}{2}$. For $V_0$ and $\alpha$ finite, $N$ is also finite. Therefore, the number of bound states is finite for the Gaussian well. 

\newpage
{\bf Acknowledgements}:\\
This work was supported by the NSF (Grants No. PHY-1912542, and
No. PHY-1607221) and the Binational (U.S.-Israel) Science Foundation (Grant No. 2015616).
Partial support from  the Department of Science and Technology(DST), India 
(under SERB core grant 
(award no. EMR/2016/005492 ) is gratefully acknowledged. 
\end{document}